\documentclass[conference]{IEEEtran}

\usepackage[utf8]{inputenc}
\usepackage[T1]{fontenc}
\usepackage{cite}
\usepackage{amsmath,amssymb}
\usepackage{graphicx}
\usepackage{booktabs}
\usepackage{tikz}
\usetikzlibrary{arrows.meta,positioning,shapes.geometric,calc,fit,backgrounds}
\usepackage{pgfplots}
\pgfplotsset{compat=1.17}
\usepackage{xcolor}
\usepackage{url}
\usepackage{balance} %

\begin{document}

\title{Measuring Bounded Autonomy for an LLM\\
Decomposition Planner on a Live 5G-Core Testbed}

\author{
\IEEEauthorblockN{Sanjeet Raj Pandey}
\IEEEauthorblockA{\textit{Service-centric Networking} \\
\textit{Technische Universit\"at Berlin}\\
Berlin, Germany \\
sanjeet.raj.pandey@tu-berlin.de %
}
\and
\IEEEauthorblockN{Axel K\"upper}
\IEEEauthorblockA{\textit{Service-centric Networking} \\
\textit{Technische Universit\"at Berlin}\\
Berlin, Germany \\
axel.kuepper@tu-berlin.de
}
}

\maketitle
\bstctlcite{BSTcontrol}

\begin{abstract}
Large language models (LLMs) are entering network-management
control loops, yet their plans can fall outside the executable action
space. Whether hard deterministic bounds around such a planner have
real effect, or merely restate a definition, is an empirical
question. We measure it on a live 5G-core testbed, where an LLM
planner decides when to decompose control-plane procedures into
stateless handlers inside a bounded loop: a JSON schema constrains
the output, a projection-repair validator re-checks it, a
deterministic fallback guarantees liveness, and every decision is
persisted. Three measurements anchor the paper.
First, the two tested models, run unbounded with the allowed
action sets stated in the prompt, still violated intent--handler
rules in 13 of 44 standard calls. Every violation broke an intent--handler combination rule that the
deployed flat schema cannot encode (schema enforcement as deployed
caught 0/13); a union schema encoding the rule removes the class by
construction. Second, with the
bounds in place, all 33 decoded live plans were feasible without
repair, and a matched-condition probe put the in-distribution
unbounded rate at one violation in 20 windows. Third, exploratory: on the design set the planner matched an
author-specified selection reference that a threshold policy and
three a priori codings each missed. Transfer to held-out scenarios
remained partial (5--6 of 10), and the planner's two-clause
distillate could replace the runtime planner itself. Together the results suggest elicitation and distillation, not
runtime control, as the LLM's role.
\end{abstract}


\section{Introduction}
\label{sec:intro}

Autonomous network management aims for self-configuring,
self-optimizing, self-healing networks \cite{ieee2021sixg,
lee2024ainetworkmanagement}; large language models (LLMs) are
entering this loop, from intent-driven operations
\cite{mekrache2025ossgpt} to agentic orchestration for 6G
\cite{jiang2026agenticibn, ameur2026agentic}.
Yet delegating management decisions to an LLM leaves open whether
outputs are valid, bounded, and auditable; trusted live deployment needs \emph{bounded autonomy}:
hard deterministic limits around a generative decision maker, each
step traceable.

Our task is adaptive decomposition of 5G core control-plane
functions: the core is already modular, with an Access and
Mobility Management Function (AMF) and a Session Management Function
(SMF)~\cite{3gpp23501}, yet deploys as always-on services,
with orchestration blind to handler-specific pressure; a
session-establishment surge stresses one SMF code path while the
rest idles. Procedure-level decomposition would let exactly the stressed path
scale; when to decompose is thus an operational decision worth
automating safely.

Function-as-a-Service (FaaS) and fine-grained network function
virtualization (NFV) are demonstrated practice in mobile cores:
serverless processing sustains 5G control-plane workloads at markedly
lower footprint~\cite{qi2022spright, gramagila2020serverless,
sabbioni2024serverless, serverless5gc2026}; function-level cores are
a stated 6G direction with AI-native
management~\cite{ieee2021sixg}, so safe automation of the decompose
decision extends beyond 5G. Establishing whether decomposition is beneficial requires a real
FaaS runtime and is future work; we address the prior question: can the decision be made safely, traceably, and cheaply enough,
with an LLM involved, for the benefit question to be posed live
at all?

Our prototype observes a live 5G core and decides when to decompose:
log-derived pressure indicators feed a deterministic policy plus a
schema-constrained LLM planner routing session events between a
baseline and a lambda-style path; the loop is
\emph{observation-closed but actuation-open}: it routes synthetic
events and never reconfigures the core; separate stub handler pods are actuated in one experiment
(Sec.~\ref{sec:livepressure}).

The mechanisms (schema, allow-list, validator, fallback) are
established; the contribution is the \emph{measurement}, on a live
system:
\begin{itemize}
  \item an \emph{unbounded-side measurement}: with the allowed
        sets stated in the prompt but unenforced, 13/44 standard
        calls violate the executable action space (0/24
        adversarial); all 13 are combination breaches the deployed
        flat schema cannot catch (0/13) and a union schema closes
        by construction (68 offline calls);
  \item a \emph{bounded-side measurement}: all 33 decoded live
        plans feasible unrepaired; a matched live probe puts the
        in-distribution unbounded rate at 1/20;
  \item an \emph{instantiation} that closes a runtime loop on live
        low-level 5G observables, formalized over a feasibility set
        (the invariant itself is definitional), every decision
        persisted, the fallback guaranteeing liveness; and
  \item a \emph{case-study evaluation}: a corrected campaign (63
        runs, 7,150 events) under live multi-SIM SMF pressure,
        quiescent runs closing the decompose/keep boundary, and ten
        handler-pod-actuated decisions.
\end{itemize}
An exploratory observation, \emph{specification elicitation}, is
reported as a hypothesis (Sec.~\ref{sec:ablation}), not a
contribution.

\section{Related Work}
\label{sec:related}

\subsubsection*{Cloud-native cores and control-plane scalability}
Cloud-native 5G design was formulated early
\cite{imadali2018cloudnative}; Kubernetes deployments of open cores
are well studied \cite{phan2024testbed, barbosa2025opensource,
barrachina2022cloudnative}, their autoscaling
\cite{yucebas2025cloudnative5gautoscaling} reactive but coarse: the
unit remains the whole function. Finer-grained approaches forecast
traffic to scale AMF instances \cite{alawe2018amfscale}, restructure
the AMF into smaller services \cite{jihang2024microamf}, or place
functions dependency-aware \cite{goshi2021nfdependecies}.

\subsubsection*{Serverless and procedure-decomposed cores}
Serverless NFV was envisioned early~\cite{aditya2019serverless} and
argued for mobile networks~\cite{gramagila2020serverless}, with
quality of service (QoS) shown preserved at the cloud
edge~\cite{sabbioni2024serverless}.
Procedure-based stateless cores externalize state per procedure (PP5GS
\cite{goshi2022pp5gs}; procedure-as-a-function private cores
\cite{serverless5gc2026}), the externalization cost quantified
in~\cite{kulkarni2021statelesscost}; CoreKube \cite{ferguson2023corekube}
and TEGRA \cite{tegra2025} add replica
autoscaling~\cite{yucebas2025cloudnative5gautoscaling} to an
already-fixed decomposition. Across this line,
granularity is a design-time decision; none treats decompose-or-not
as an online control variable.

\subsubsection*{AI- and LLM-based network management}
Surveys document AI's shift from external tool to architectural
component~\cite{lee2024ainetworkmanagement, dubey2024airesoucemanagement,
ieee2021sixg}; LLMs are adapted to networking~\cite{wu2024netllm} and
telecom operations~\cite{maatouk2024llmtelecom}. A fast-growing line
centers LLMs in autonomous operations:
LLMs extract 3GPP intents \cite{manias2024intent}, benchmark
configuration synthesis \cite{netconfeval2024}, front intent-driven
operations \cite{mekrache2025ossgpt}, and translate intents to
network domain-specific languages (DSLs) with parser/dry-run
validation (Confucius \cite{confucius2025}) or coordinate RAN and
core agents \cite{jiang2026agenticibn, ameur2026agentic}. The closest design
guards a 6G-RAN command DSL with an allow-list validator and a
safety-filter adapter (preprint)~\cite{ranoperator2025}. This framing
draws on established mechanisms: programmable guardrails
\cite{rebedea2023nemo}, generate-then-verify ``LLM-modulo'' planning
\cite{kambhampati2024llmmodulo}, and shielding
\cite{alshiekh2018shielding}. These systems operate mostly at the
offline intent/config-authoring level with human-in-the-loop gates
(Table~\ref{tab:mech}).

\subsubsection*{Positioning}
We complement these intent-level agentic systems one level down
(Table~\ref{tab:mech}); we are not aware of a prior prototype closing a
bounded runtime loop on live low-level observables of an open 5G core.

\begin{table}[t]
\caption{Mechanisms as described in each paper (not artifacts we
ran; \cite{ranoperator2025} is a vision preprint; -- = absent or not
stated).}
\label{tab:mech}
\centering\scriptsize
\begin{tabular}{@{}lccc@{}}
\toprule
Mechanism & \cite{ranoperator2025} & Confucius~\cite{confucius2025} & this work \\
\midrule
Output schema / DSL constraint & \checkmark & \checkmark & \checkmark \\
Semantic validator (allow-list) & \checkmark & \checkmark & \checkmark \\
Deterministic fallback (liveness) & -- & -- & \checkmark \\
Runtime loop on live observables & -- & -- & \checkmark \\
Runs without a human gate & -- & -- & \checkmark \\
Per-decision persistence (audit) & -- & -- & \checkmark \\
\bottomrule
\end{tabular}
\end{table}

\section{System Design}
\label{sec:design}

\begin{figure}[t]
\centering
\resizebox{0.56\columnwidth}{!}{%
\begin{tikzpicture}[
  font=\scriptsize,
  box/.style={draw, rounded corners=1pt, align=center, fill=white,
              inner sep=2pt, minimum height=6mm, minimum width=19mm},
  gate/.style={draw, diamond, aspect=1.8, align=center, fill=white,
               inner sep=1pt, minimum height=6mm},
  lay/.style={draw, dashed, rounded corners=3pt, inner sep=3.5pt},
  laylab/.style={font=\tiny\bfseries, rotate=90, anchor=south, inner sep=1pt},
  arr/.style={-{Stealth[length=2mm]}, semithick},
  fb/.style={-{Stealth[length=2mm]}, semithick, densely dashed}
]
\node[box] (ueransim) {UERANSIM\\UE/gNB workload};
\node[box, right=8mm of ueransim] (free5gc) {free5GC core\\AMF / SMF pods};
\draw[arr] (ueransim) -- node[above]{traffic} (free5gc);
\node[box, below=5mm of ueransim] (metrics) {pressure metrics\\$P_{\mathrm{SMF}},P_{\mathrm{AMF}},P_{\mathrm{we}}$};
\node[box, right=8mm of metrics] (parser) {AMF / SMF\\log parsers};
\draw[arr] (free5gc.south) -- node[right,pos=0.45]{logs} (parser.north);
\draw[arr] (parser) -- (metrics);
\node[box, below=5mm of parser] (policy) {policy engine\\threshold guardrail};
\node[box, right=8mm of policy] (llm) {LLM planner\\schema-bounded};
\draw[arr] (metrics) -- (policy);
\draw[arr] (policy) -- node[above]{mode} (llm);
\node[gate, below=5mm of $(policy)!0.5!(llm)$] (val) {validator\\$\hat p\!\in\!\mathcal F$?};
\draw[arr] (llm) -- (val);
\node[box, left=8mm of val, fill=black!5] (fb) {fallback\\$f(x)\in\mathcal F$};
\draw[arr] (val) -- node[above]{no} (fb);
\node[box, below=5mm of val] (dispatch) {controller $+$ dispatcher};
\draw[arr] (val) -- node[right]{yes} (dispatch);
\draw[arr] (fb.south) |- ([yshift=1.5mm]dispatch.west);
\node[box, below left=5mm and -8mm of dispatch] (baseline) {baseline\\path};
\node[box, below right=5mm and -8mm of dispatch] (lambda) {lambda-style\\handlers};
\draw[arr] (dispatch) -- (baseline);
\draw[arr] (dispatch) -- (lambda);
\node[box, below=4mm of $(baseline.south)!0.5!(lambda.south)$] (store) {state store $\cdot$ CSV trace};
\draw[arr] (baseline) -- (store);
\draw[arr] (lambda) -- (store);
\draw[fb] (store.east) -- ++(14mm,0) |- node[pos=0.25,right,align=left]{activation\\(next cycle)} (dispatch.east);
\begin{scope}[on background layer]
\node[lay, fit=(ueransim)(free5gc), label={[laylab]left:Environment}] {};
\node[lay, fit=(parser)(metrics), label={[laylab]left:Observation}] {};
\node[lay, fit=(policy)(llm)(val)(fb), label={[laylab]left:Decision}] {};
\node[lay, fit=(dispatch)(baseline)(lambda)(store), label={[laylab]left:Execution}] {};
\end{scope}
\end{tikzpicture}}
\caption{Architecture: a loop closed at observation (the core is read,
not reconfigured); the validator repairs decodable plans by projection into
$\mathcal F$ and substitutes the deterministic fallback only for
undecodable output.}
\label{fig:architecture}
\end{figure}
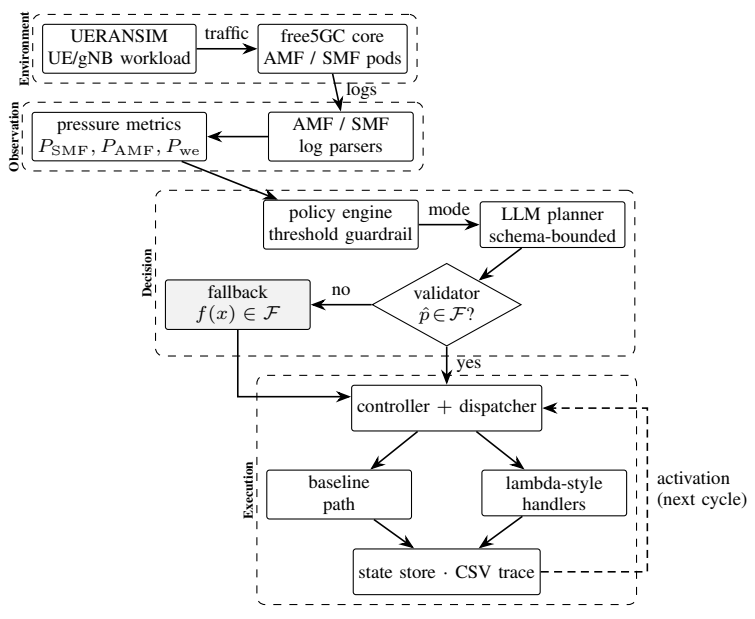

Fig.~\ref{fig:architecture} shows the loop in four layers: the observer snapshots AMF/SMF logs around a workload
burst; parsers compute delta counters and three pressure indicators;
the policy engine picks a mode; the planner proposes a handler set;
the validator repairs or substitutes;
the controller activates permitted handlers; the dispatcher routes
each event via the baseline or lambda path.

\subsection{Observation Layer}
A Kubernetes observer snapshots the AMF and SMF pod logs before and after
a workload burst; parsers extract \emph{delta counters}, the
per-event-type activity counts occurring between the two snapshots:
SMF-side session create/modify/release activity, Protocol Data Unit
(PDU) session reports, Packet Forwarding Control Protocol
(PFCP)/N4 activity, active sessions,
warning/error counts; AMF-side registration, authentication, and
context events with warning/error/retry indicators. The deltas are
condensed into three scalar pressure indicators (formalized in
Sec.~\ref{sec:formal}): SMF pressure over session and weighted
N4/PFCP activity, AMF pressure over registration and authentication
signals, and warning/error pressure; with the runtime metrics they
form the decision input.

\subsection{Two-Stage Decision Layer}
Decision making is split: first, a deterministic
policy engine maps the metrics to one of four policy decisions
(\texttt{baseline\_mode}, \texttt{lambda\_mode}, \texttt{scale\_up},
\texttt{scale\_down}) using explicit threshold pairs $X/Y$ per
metric: crossing $X$ selects \texttt{lambda\_mode}, crossing $Y$
\texttt{scale\_up} (defaults SMF 30/90, warn/err 10/30, AMF 20/60;
live campaigns tune SMF to 8/20 and disable warn/err); the decision gates how wide a
handler set the planner may activate. This stage is the primary guardrail:
auditable, reproducible, unable to hallucinate. Second, an LLM planner receives the same structured metrics plus the
policy decision and produces a decomposition plan, bounded three ways:
(i)~its output must fit a JSON schema whose intent, mode, and handler
fields are \texttt{enum}-constrained to an allow-list, as in guided
decoding~\cite{willard2023guided} (full schema in the artifact); (ii)~a deterministic validator
re-checks intents and repairs decodable plans by \emph{projection},
dropping handlers outside $\phi(\iota)$; (iii)~the
controller activates only allow-listed handlers.
Only when no plan can be
decoded (unparseable output or model unavailability) does a
deterministic fallback substitute ($f(x)$: every handler admissible
under the policy mode, blanket coverage trading efficiency for
availability), so the loop never blocks on the model; every step is persisted for
audit.

\subsection{Formal Model and Safety Invariant}
\label{sec:formal}
The pressure indicators are weighted sums of the log-derived deltas:
$P_{\mathrm{SMF}}=\sum_i w^{\mathrm{s}}_i\,\delta^{\mathrm{s}}_i$,
$P_{\mathrm{AMF}}=\sum_i w^{\mathrm{a}}_i\,\delta^{\mathrm{a}}_i$, and
$P_{\mathrm{we}}$ the summed AMF/SMF warnings and errors (disabled in
the tuned live configuration); the policy thresholds these three
scalars (weights and values in the artifact).
Let $H$, $I$, $M$ be the finite sets of allowed handlers, intents, and
modes, and $\phi\!:\!I\to 2^{H}$ the handlers admissible for each intent. A
plan $p=(\iota,\mu,S)$, with intent $\iota$, mode $\mu$, and selected
handlers $S\subseteq H$, is \emph{feasible} iff $\iota\in I$, $\mu\in M$, and
$S\subseteq\phi(\iota)$; let $\mathcal{F}$ be this feasible set.
Concretely $|H|{=}4$, $|M|{=}4$, $|I|{=}8$, $|\mathcal F|{=}208$;
$\phi$: stable/recovery${\to}\emptyset$;
registration surge${\to}$\{obs\}; establishment
surge${\to}$\{obs, create\}; release surge${\to}$\{obs, release\};
timeout/error${\to}$\{create, modify, release\}; degradation and
scale-up${\to}$all four (obs = \texttt{registration\_observer}). The LLM maps the
metric state $x$ to raw output $o$; $\hat{p}(o)$ decodes a candidate
plan. The bounded planner applies a
validator $V$ with projection repair
$\pi(\iota,\mu,S)=(\iota,\mu,\,S\cap\phi(\iota))$ (identity on
feasible plans; out-of-enum $\iota,\mu$ count as undecodable) and a
deterministic fallback $f(x)\in\mathcal{F}$,
\begin{equation}
\label{eq:invariant}
V(o,x)=\begin{cases} \pi(\hat{p}(o)) & \hat{p}(o)\ \text{decodable}\\ f(x) & \text{otherwise,}\end{cases}
\end{equation}
so $V(o,x)\in\mathcal{F}$ for every $o$: the activated plan is
in-scope by construction (a one-step shield~\cite{alshiekh2018shielding}).
The guardrail-off study (Sec.~\ref{sec:containment}, RQ3) reports
$\Pr[\hat{p}(o)\notin\mathcal{F}]$: the violation mass this invariant
removes. Input integrity is a stated non-goal (residual measured in
Sec.~\ref{sec:discussion}).

\subsection{Lambda-Style Execution and State}
Supported session events (\texttt{session\_create},
\texttt{session\_modify}, \texttt{session\_release}) can be executed
either on the baseline path (conventional in-process handling) or by
stateless, lambda-style handlers (in this prototype:
decomposition-path stubs sharing the process). Decomposition moves state out of
the function rather than eliminating it: an external store keeps
session context (identifier, user-plane function, QoS, slice, state)
consistent across handler invocations; handlers lock, update, and
release records per event; a dispatcher routes by mode, logging
outcomes.

\section{Prototype Implementation}
\label{sec:impl}

The prototype is Python on free5GC~\cite{free5gc} v3.3.0 on
Kubernetes (k3s); UERANSIM~\cite{ueransim} (v3.2.6) provides
next generation Node B (gNB) and
user equipment (UE) registration bursts (UE scale-out to five per
burst); the
observer snapshots pod logs via the Kubernetes API around each burst. The LLM planner calls \texttt{gpt-5.4-mini} (snapshot
2026-03-17, temperature 0.0) with structured JSON-schema output; the
prompt embeds the allowed intents, modes, and handlers, and a lightweight
history mechanism feeds prior-run summaries back as context
(disabled in the ablation to avoid confounding); the full prompt,
schema, configuration, and ablation harness ship with the released
source. The workload generator persists seed-generated
session events for identical replay across modes; an experiment
runner writes the per-event and per-decision CSVs behind all
results.\footnote{A replication package (harness, testbed scripts,
datasets, analysis code) is available at
\url{https://github.com/sanjugem/anms2026-artifact}.}

\section{Evaluation}
\label{sec:eval}

\subsection{Setup and Method}
The primary evidence is the corrected campaign: 63 runs with
shortened replays (about 113 events per run) of two fixed, seed-generated workloads
(creation-heavy and balanced), run under session churn (continuous
create/release cycling) unless noted. Each run executes in one of three
modes: baseline-only, lambda-only, or AI-controlled, in which the
policy/LLM layer decides routing after a live free5GC observation.
Table~\ref{tab:campaigns} maps all campaigns, including the superseded 90-run original. Latencies are in-process
figures, indicative only (Sec.~\ref{sec:intro}). We ask: (RQ1) are decisions
traceable and the loop live under model failure? (RQ2) is the
bounded planner's selectivity consistent with a stated reference,
and robust across models and unseen inputs? (RQ3) how much
out-of-scope behavior does an unbounded planner produce, and does
the bounding contain it? Overhead is negligible: inference is off-path.\footnote{The superseded campaign's 27,000
executions all succeeded (100\%); this is structurally expected, as both
paths share one process; its 675 schema-valid decisions (all from its 30 AI-controlled runs,
about 22 per run, one per observation cycle within the long
replays) corroborate throughput only.}

\begin{table}[t]
\caption{Claim-to-campaign map: every quantitative claim traces to
one row (scale and condition in col.~2).}
\label{tab:campaigns}
\centering\scriptsize
\begin{tabular}{@{}p{1.05cm}p{2.55cm}p{3.6cm}@{}}
\toprule
Campaign & Scale / condition & Supports \\
\midrule
Original & 90 runs, 27k exec.; 1 IMSI, SMF $=$ 0 (\textbf{defect}), default thr. & superseded; 675 schema-valid decisions \\
Corrected & 63 runs (15/15/33 base/lambda/AI), shortened replays, 7,150 ev.; churn, tuned thr. & 34 live decisions; narrowing 30/33 \\
Quiescent & 10 runs; churn off & keep side: \texttt{baseline\_mode} 10/10 \\
Matched & 20 windows, churn on; history on & unbounded 1/20; fallbacks 0/20 \\
Union schema & 68 offline calls, strict flat vs.\ union & 0 breaches both; selections identical \\
Actuated & 10 decisions; churn on/off & pod actuation, rollback \\
\bottomrule
\end{tabular}
\end{table}

\subsection{Traceability and Liveness (RQ1)}
RQ1 rests on the corrected campaign's 34 fully persisted
decisions. Self-reported confidence does not discriminate: across the 80 held-out replay decisions (Sec.~\ref{sec:ablation};
10 scenarios $\times$ 4 backing models $\times$ 2 reps; 42 matching
the reference, 38 not),
mean confidence differs by only 0.017; it is persisted for audit
only. Every decision is fully reconstructible from the persisted
artifacts (metrics, policy decision, raw response, validation,
activated handlers).

\subsection{Planner Selectivity and Generalization (RQ2)}
\label{sec:ablation}
\begin{figure}[t]
\centering
\includegraphics[width=0.60\columnwidth]{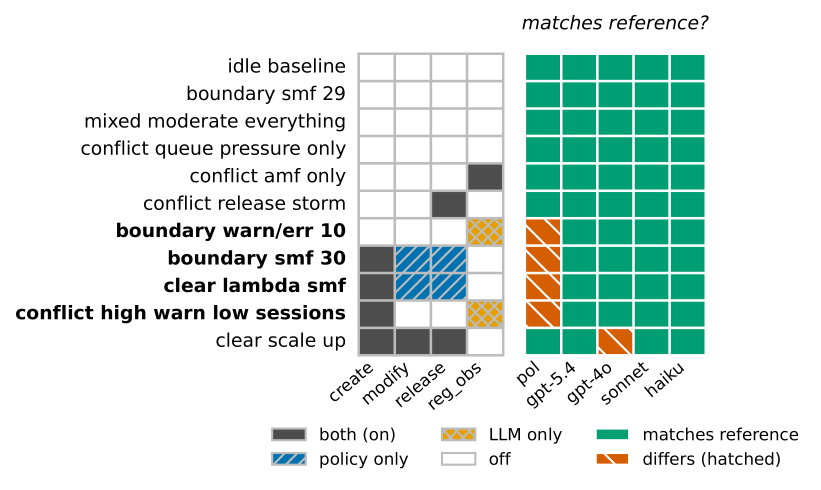}
\caption{Left matrix: which of the four handlers the deterministic
policy vs.\ the LLM planner activates on each of the eleven ablation
scenarios (identical inputs, default thresholds); bold rows are
the four scenarios where they diverge. Right strip: each column scores
one system's handler set against the frozen a priori reference
(green = exact match, hatched = differs): three models 11/11,
\texttt{gpt-4o-mini} 10/11, the policy 7/11.}
\label{fig:ablation}
\end{figure}

To isolate what the LLM adds beyond the threshold policy, we replayed
eleven synthetic metric scenarios spanning clear-cut overloads,
threshold-boundary cases, and conflicting signals (full metric
vectors ship with the artifact) against the planner and the deterministic fallback
on identical inputs, five times each (55 calls;
\texttt{gpt-5.4-mini}); the policy runs its default thresholds, the
charitable choice, because the tuned live thresholds match the
a priori selection reference (defined below) on only 6 of the 21
evaluation scenarios (11 design $+$ 10 held-out). Live, the tuned policy never holds this selection role; it
only gates mode width. All 55 responses were schema-valid, allow-listed,
mode-consistent, and identical across repetitions (temperature 0.0; provider determinism
is not guaranteed, so stability is an observation, not a property). The planner diverged from the fallback in 20/55 queries (36\%),
consistently (four scenarios in all five repeats, the rest in none):
where the fallback blanket-activates, the planner restricts
decomposition to the handler backed by the dominant metric, adding
\texttt{registration\_observer} under AMF warning/error pressure
(Fig.~\ref{fig:ablation}).
The planner is invoked once per observation window; median latency
was 2.97\,s per call vs.\ a median 40\,s between windows
(645 intervals; a testbed cadence), so each decision completes
before the next observation; at sub-second control timescales this
architecture is inapplicable. The planner is confined to slower
reconfiguration decisions, with deterministic stages covering fast
paths. Model unavailability was exercised live once
(Sec.~\ref{sec:livepressure}): the call fell back without blocking
the loop.

To test whether this selectivity is consistent, we scored each scenario against an author-specified
a priori reference (offload only the procedure under load; observe
registration when signaling rises); ``correct'' throughout means
consistent with this rule (Sec.~\ref{sec:discussion}); labels were
frozen before scoring and ship with the artifact. The planner matched it in all eleven scenarios and all 20
divergent decisions; the policy matched seven, over-decomposing in
two divergences, missing the observer in two.

\emph{Robustness and generalization (RQ2).} Replaying all eleven scenarios through four backing models
(\texttt{gpt-5.4-mini}, \texttt{gpt-4o-mini}, \texttt{claude-sonnet-4-5},
\texttt{claude-haiku-4-5}; two families) under identical schema and
prompt (two repetitions each), all 88 replays were schema-valid and
stable across them; three models match the reference 11/11 with identical sets,
\texttt{gpt-4o-mini} 10/11. To test transfer, we built ten held-out
patterns absent from the design set (modify-dominant load, teardown
waves, signaling storms, surges with registration errors)
and scored all four models and the frozen policy against the same
reference. The result is largely negative: the models matched the reference
only 5--6/10 on unseen inputs; selectivity does not transfer
reliably, though the errors were structured (omissions of
a co-dominant procedure, never spurious activations, 0--1/10 idle
activations vs.\ the frozen policy's 4/10; the pattern held
across models).

Three a priori codings of the rule's sentence score 9/11 each on
the design set (each missing a different pair, where the planner
scores 11/11) and 14--17/21 overall vs.\ the models'
15--17/21, overlapping at n=21; each coding fails on a different
choice the sentence leaves open. The planner's
divergence, distilled post hoc into a two-clause rule (maximal
positive session-delta handler(s) under a lambda/scale-up gate;
observer when signaling crosses threshold), scores 7/10 held-out
vs.\ the models' 5--6/10 (10/11 on its own design set,
contaminated by construction); the encoded divergence is identical
across the three 11/11 models. The backing models' measured
contribution is therefore not selection accuracy but
\emph{specification elicitation}: deriving zero-shot what each a priori
coding got wrong in a different way; the distillate then runs free of
cost and latency (a hypothesis at these counts).

\subsection{Guardrail-Off and Matched-Condition Containment (RQ3)}
\label{sec:containment}
To quantify what the bounding contributes, we re-ran the eleven
scenarios and six adversarial inputs unbounded (no schema, validator,
or fallback) on \texttt{gpt-5.4-mini} and \texttt{claude-sonnet-4-5}
(n=22 standard and n=12 adversarial calls per model, i.e.,
44 and 24 in total). With the allowed action
space omitted from the prompt, 82--100\% of plans fell outside it
(invented handlers, out-of-enum modes; 100\% adversarial); expected
when the enums are unseen, reported only as the floor. The informative condition keeps the allowed sets in the prompt: 13
of 44 standard calls still broke the intent--handler rules the
validator enforces (5/22 and 8/22 per model, exact 95\%
intervals 8--45\% and 17--59\%; the pooled 13/44 is descriptive,
aggregating two models with different rates; 0/24
adversarial); prompt-stated rules alone do not contain the
backing model. Violations concentrated on the same three boundary scenarios in
both models, suggesting input-driven rather than
model-idiosyncratic behavior. This inversion (adversarial inputs clean, standard inputs
violating) is informative: extreme inputs appear to be treated as
anomalous, drawing conservative, often empty, sets within $\phi(\iota)$,
while plausible boundary inputs invited over-selection across
intent--handler lines. All 13 informed-condition violations were intent--handler
\emph{combination} breaches ($S\not\subseteq\phi(\iota)$); the deployed flat
schema of per-field \texttt{enum}s cannot express that rule, so
schema enforcement as deployed caught 0/13. A discriminated-union schema (one
branch per intent, handler enums restricted to $\phi(\iota)$) encodes
the rule; replayed offline on the same 17 scenarios under strict
constrained decoding (68 calls), it excluded the breach class by
construction and matched the strict-decoded flat schema call for
call (34 pairs). Even the flat schema, decoded strictly, produced no breaches where
free-form output violated: the union form adds a guarantee, not an
observed rate difference; the validator slot remains responsible
for stateful and impact-aware rules. Projection, the deployed repair
(Sec.~\ref{sec:formal}), scored post hoc on the 13 recorded raw
violations preserves reference agreement 8/13 vs.\ 5/13 for blanket
substitution (within noise at n=13). Under the full schema and validator the violation rate is 0\%,
realizing the
containment of Eq.~\eqref{eq:invariant} (RQ3): the schema blocks
invented actions, the validator unpermitted ones. A matched-condition live probe (20 churn-on windows;
\texttt{gpt-5.4-mini}; history on, tuned thresholds; observed SMF pressure 1--50;
paired unbounded/bounded calls per window) reconciles this with the
zero live
violations (Sec.~\ref{sec:livepressure}): in distribution the
unbounded rate falls to 1/20 (95\% interval 0--25\%), but not to
zero. The survivor is an intent--handler combination breach;
executed, it would mean spurious decomposition (costless on stubs,
spurious scale-out on a real runtime); the distillate cannot commit this class of breach, having no
intent field to mismatch (at the cost of its audit semantics).

\subsection{Live SMF Pressure: Multi-SIM Runs}
\label{sec:livepressure}
The original campaign carried a provisioning defect, reported openly:
with one subscriber every PDU establishment was rejected (5GSM cause~\#50), leaving SMF pressure zero and decisions
driven by AMF warning/error pressure. We corrected the data, provisioned ten
subscribers (International Mobile Subscriber Identities, IMSIs),
each verified with registration and a real PDU session (10/10),
and re-ran under sustained session cycling (Table~\ref{tab:campaigns}; 100\%
lifecycle success). The corrected campaign's 33 AI runs yielded 34 decisions: 33
live-LLM plus one availability-triggered fallback (a second
observation window straddled a run boundary; its call found the API
key unset), all under nonzero live SMF pressure (2--44, median 18). Handler counts cover the 33 live-LLM decisions; substitution counts cover all 34. The policy exercised \texttt{lambda\_mode} and \texttt{scale\_up}
(never \texttt{scale\_down} live, a coverage gap); intents spanned establishment/release surges, degradation, and
scale-up. The planner stayed selective
on real load: 30 of 33 live decisions narrowed the handler set with zero raw violations (history is
enabled live but not in the ablation, so live selectivity is not
history-free). The bounding's measured price was zero validator-triggered
substitutions in 34 decisions; the single substitution was that
availability-triggered call, not a response to infeasible output.
Churn is sustained deliberately, so \texttt{baseline\_mode} is out
of reach there; ten quiescent runs (churn paused) close the gap:
\texttt{baseline\_mode} 10/10, planner
concurring with no handlers, exercising the boundary from both
sides. The corrected and quiescent campaigns
run the artifact's tuned thresholds, recalibrated from the earlier
sweep before the evaluated runs, not on them (warn/err disabled because
the UE scale-out burst (Sec.~\ref{sec:impl}) injects a warning
floor).

\emph{Actuated decisions.} Ten live decisions (five under churn, five
quiescent; disjoint from the 34) realized their validated handler
sets as Kubernetes scale operations on four handler deployments behind an
actuation allow-list mirroring the planner's. All ten plans
(observer-only to full session set; SMF 0--641, the peak reflecting
the scale burst superimposed on churn)
converged in 0.6--4.2\,s, rollback to zero verified per batch
via the API server; this verifies the actuation channel only, since
the pods are stubs: actuation, not traffic.

\section{Discussion and Limitations}
\label{sec:discussion}

\subsubsection*{From routing to actuation}
The campaigns route synthetic events; ten decisions actuate handler-pod
lifecycles (Sec.~\ref{sec:livepressure}); the core is never
reconfigured. Core-touching actuation raises the stakes: feasibility-set
membership is necessary but not sufficient; an in-scope plan can
still be the wrong action, so such deployments need impact-aware
checks in the validator slot. Whether containment scales beyond $|\mathcal F|{=}208$ is open:
larger, stateful spaces grow both the violation surface and the
validator's job; the structural lesson (combination constraints
escape flat schemas) should transfer; the rates need not. A sequence-aware $\mathcal F$
would constrain trajectories rather than steps, with dwell times, hysteresis
margins, and rate limits as validator state: $V$ becomes stateful. The elicitation result (Sec.~\ref{sec:ablation}) suggests a deployment
pattern: elicit with the bounded planner, distill, deploy the
rule; the runtime controller is then deterministic, the planner
re-invoked only on drift (the detector is future work). This repositions the LLM from runtime controller to offline rule
author, grounded in runtime observables.

\subsubsection*{Prototype scope and limitations}
The handlers execute in-process, not on a FaaS runtime: no cold
starts, no contention to relieve; latencies are prototype-internal.
We establish only that the decision can be made safely, traceably, and
cheaply; its resource benefit is future work. The session lifecycle model (create/modify/release) simplifies
the full PDU-session procedure space~\cite{3gpp23502}; the state
store is in-memory, not distributed. Scaling the corrected campaign
to the original 27,000-execution volume remains future work, as do
broader workloads. We scope out learned controllers: no reward signal yet justifies
them.

\subsubsection*{Threats to validity}
Decisions rely on log parsing (format changes affect the metrics);
results are from one host and core. The ablation is scored against an author-specified a priori rule,
corroborated by cross-family agreement, not externally
standardized; a different rule could label the
conflicting scenarios differently, and counts are small (n=11,
n=10), so proportions are counts, not significance claims.
The invariant is per-decision, and mode flapping occurs: the
matched probe's 20 live windows switched lambda/scale-up modes in
10/19 consecutive pairs (the decompose/keep boundary never
crossed); replaying the windows through a two-window dwell rule suppresses
9/10 switches; minimal hysteresis, encoded as validator state,
would likely tame it; live enforcement is a listed next
experiment. Exact intervals are inline and in the artifact.
The validator bounds outputs, not inputs; we measured the residual: ten
corrupted-telemetry scenarios (negative counters, physically impossible magnitudes,
inconsistent combinations; three repetitions) yielded 30/30
schema-valid plans, yet six induced handler activation, valid but operationally
wrong. Six a priori range/consistency rules detect
10/10 with zero false positives (21 scenarios, 223 live
windows), though corruptions and rules share authorship; wiring the gate
in is future work. The study is numeric; adversarial log \emph{text} cannot reach the
prompt (counters only), though it could distort parsing (untested). Pinning a commercial
snapshot risks deprecation and silent change; deprecation degrades
to the fallback rather than blocking, while
cross-model results show the bounding survives a swap (selectivity
needs re-validation). The prompt carries only aggregate,
identifier-free counters.

\section{Conclusion and Future Work}
\label{sec:conclusion}

This paper measured bounded autonomy for an LLM planner on a live
5G-core testbed: log-derived pressure metrics drive a two-stage
decision layer routing session events between baseline and lambda
paths. The bounding is a validator over a feasible
action set: with it removed, 13 of 44 standard calls violate that
set
(1/20 live, in distribution); across four
models the planner diverged selectively from the policy, matching,
on the design set, an author-specified reference the policy only
partially meets; the distillate of that divergence is a candidate
replacement, an exploratory signal.

The discriminated-union condition has since been run
(Sec.~\ref{sec:containment}); one next experiment runs on the
released harness as-is: live enforcement of the two-window dwell
rule. Further out: handlers on a real
FaaS runtime, scaling the corrected campaign, an open-weights
on-premises planner, independent reference labeling.

\section*{Acknowledgment}
The authors thank Ivaylo Daskalov for the early testbed prototype
from his master's thesis, Diane Wagner for proofreading, and the
anonymous reviewers for their constructive feedback.
The text was in part
drafted and edited with LLM assistance under the
authors' direction; all experiments, data, and claims were designed,
executed, and verified by the authors, who take full responsibility
for the content.
\balance
\bibliographystyle{IEEEtran}
\bibliography{references}

\end{document}